\documentclass[12pt]{article}
\usepackage{amssymb}
\usepackage{amsfonts}
\usepackage{amsmath}
\usepackage{mathrsfs}

\def\sb{{\sf b}}
\def\cH{{\cal H}}
\def\cK{{\cal K}}
\def\tr{{\rm tr}}

\newcommand\ket[1]{| #1 \rangle}

\newcommand\braket[2]{\langle #1|#2\rangle}
\newcommand\ketbra[2]{|#1\rangle\langle #2|}

\newcommand\trnorm[1]{\| #1 \|_1}

\begin{document}

\title{{\bf Quantum bit
    commitment\\
and unconditional security}}

\author{{\bf Horace P. Yuen}\thanks{E-mail:  {\tt
      yuen@ece.northwestern.edu}}
\thanks{
{\bf Note:} this paper analyzes in detail, for the first
time, the various gaps in the QBC impossibility proof, many of which I
indicated before but few of which seem to be understood.  There
is clearly a need to focus on these gaps, which is an issue logically
distinct from whether any protocol can be proved unconditionally
secure.  One of the original three protocols I
described in the QCM at Capri, July 2000, protocol {\bf Y3} that
      appeared as {\bf QBC1} in v2 of this paper, was pointed out to be
      insecure in the QCM at MIT, July 2002.  It is
      extended, with the same underlying idea, and renamed QBC4 in
      this v3 with a full security proof.  A considerable amount of
      new material is also added, including a protocol based on
      cheating detection alone and clarifications on unknown
      parameters and entanglement purification.  It is also stressed
      that a priori there can be no impossibility proof without
      a QBC definition.}
\\[12pt]
Department of Electrical and Computer Engineering\\
Department of Physics and Astronomy\\
Northwestern University\\
Evanston, IL 60208-3118}

\date{}
\maketitle

{\abstract

\noindent{It is generally believed that
  unconditionally secure quantum bit commitment is impossible, due to
  widespread acceptance of an impossibility proof that utilizes
  quantum entaglement cheating.  In this paper, we delineate how the
  impossibiliy proof formulation misses various types of quantum bit
  commitment protocols based on two-way quantum communications.  We
  point out some of the gaps in the impossibility proof reasoning, and
  present corresponding
  counterexamples. Four different types of bit
  commitment protocols are constructed with several new protocol
  techniques. A specific Type 4 protocol is described and proved
  unconditionally secure. Security analysis of a Type 1 protocol and
a Type 2
protocol are also sketched.  The security of Type 3 protocols is as
yet open. A development of quantum statistical decision theory and
quantum games is needed to provie a complete security analysis of many
such protocols.}}

\newpage

\tableofcontents
\newpage

\baselineskip 24pt

\section{Introduction}\label{sec:intro}

There is a nearly universal acceptance of the general impossibility of
secure quantum bit commitment (QBC), taken to be a consequence of the
Einstein-Podolsky-Rosen (EPR) type entanglement cheating which
supposedly  rules out QBC and other quantum protocols that have been
proposed for various cryptographic objectives \cite{bc}.  In a {\it bit commitment} scheme, one party, Adam, provides another party,
Babe, with a piece of evidence that he has chosen a bit \sb\ (0 or 1)
which is committed to her.  Later, Adam would {\it open} the commitment
by revealing the bit \sb\ to Babe and convincing her that it is indeed
the committed bit with the evidence in her possession, which she can
{\it verify}.  The usual concrete example is for Adam to write down the bit
on a piece of paper, which is then locked in a safe to be given to
Babe, while keeping for himself the safe key that can be presented
later to open the commitment.  The evidence should be {\it binding},
i.e., Adam should not be able to change it, and hence the bit, after
it is given to Babe.  It should also be {\it concealing}, i.e., Babe
should not be able to tell from it what the bit \sb\ is.  Otherwise,
either Adam or Babe would be able to cheat successfully.

In standard cryptography, secure bit commitment is to be achieved
either through a trusted third party, or by invoking an unproved
assumption concerning the complexity of certain computational
problems.  By utilizing quantum effects, specifically the intrinsic
uncertainty of a quantum state, various QBC schemes not
involving a third party have been proposed to be unconditionally
secure, in the sense that neither Adam nor Babe could cheat with any
significant probability of success as a matter of physical laws.  In
1995-1996, a supposedly general proof of the impossibility of
unconditionally secure QBC, and the insecurity of previously proposed
protocols, was presented \cite{may1}-\cite{lc3}.  Henceforth it has
been generally accepted that secure QBC and related objectives are
impossible as a matter of principle \cite{lo}-\cite{mqi}.

There is basically just one impossibility proof, which gives the EPR
attacks for the cases of equal and unequal density operators that Babe
has for the two different bit values.  The proof purports to show that
if Babe's cussessful cheating probability $P^B_c$ is close to the
value 1/2, which is obtainable from pure guessing of the bit value,
then Adam's successful cheating probability $P^A_c$ is close to the
perfect value 1.  This result is stronger than the mere impossibility
of unconditional security, namely that it is impossible to have both
$P^B_c \sim 1/2$ and $P^A_c \sim 0$.  The impossibility proof
describes the EPR attack on a specific type of protocols, and then
argues that all possible QBC protocols are of this type.

Typically, one would expect that a proof of impossibility of carrying
out some thing X would show that any possible way of doing X would
entail a feature that is logically contradictory to given principles,
as, for example, in the cases of quantum no-cloning \cite{wz,yue1} and
von Neumann's no-hidden-variable theorem \cite{neu}.  In the present
case, one may expect a proof which shows, e.g., that {\it any} QBC
protocol that is concealing is necessarily not binding. It is
important for this purpose that the framework of QBC protocol
formulation is all-inclusive.  In the
absence of a proof that all possible QBC protocols have been included
in its formulation, any impossibility proof is at best incomplete.
Indeed, in the QBC impossibility proof, only certain
techniques of protocol design, such as the use of classical random
numbers in a quantum protocol, are included in its formulation without
showing that all possible techniques have been included.  In this
paper, we will describe several new techniques that are not
accounted for in the impossibility proof formulation.

There are two related assertions in the impossibility proof that are
crucial to both its claim of universality in general, and its specific
claim of covering the use of random numbers in particular.  These are
the assertions that all measurements in the commitment phase of a
quantum protocol can be postponed until the opening and the
verification phases, and that classical random numbers can be
equivalently described by pure quantum states, via quantum
purification or the doctrine of ``Church of the Larger Hilbert
Space.''  In this paper, we will extensively analyze the serious
problems associated with these assertions.

The essential argument of the general impossiblity proof is described
in Section \ref{sec:impproof}, and some of its problems are indicated in Section
\ref{sec:probs}. A proper framework for QBC
protocols is discussed in Section \ref{sec:proper}.  In
Section \ref{sec:newprot}, we describe several new protocol techniques
that lead to the development of four new types of protocols not
covered by the impossibility proof. In Section \ref{sec:nopost} we
describe Type 1 protocols, in which the postponement of a measurement
until opening and verification would yield a protocol with different
cheating performance.  A specific protocol {\bf QBC1} is presented with a
sketch of the security proof. In Section \ref{sec:space}, the
logic underlying Type 2 protocols is delineated.  A specific protocol,
{\bf QBC2A}, is presented with an outline of the security proof. The
security analysis of protocols {\bf QBC1} and {\bf QBC2A} are not
complete in the sense that exact optimality can only be proved with a
sequential quantum decision theory yet to be developed, althought all
essential points are included under the assumption that no party can
cheat if it can be detected with a nonvanishing probability before a
bit is committed.  In Section \ref{sec:random}, the
widely accepted equivalence between classical randomness and quantum
purification is analyzed.  We will show that they are not equivalent in bit
commitment. We also introduce Type 3 protocols, the security status of
which is yet undecided. In Section \ref{sec:type4}, we introduce Type
4 protocols which involve Babe's open questions related to Adam's
committed evidence. A specific protocol {\bf QBC4} is proved
unconditionally secure. The last Section \ref{sec:concl} contains a brief summary of the
main points.  The appendices, especially Appendix B, are an integral
part of the paper, being separated for convenient organization of this
rather sutble and multi-faceted subject.  Also, the different types of
protocols in this paper are not mutually exclusive.  Again, they are mainly
introduced for the purposes of organization.

\newpage

\section{The impossibility proof:  Type 0 protocols}\label{sec:impproof}

The impossibility proof, in its claimed generality, has never been
systematically spelled out in one place, but the essential ideas that
constitute this proof are generally agreed upon
\cite{may2}-\cite{mqi}.  The formulation and the proof can be cast as
follows.  Adam and Babe have available to them two-way quantum
communications that terminate in a finite number of exchanges, during
which either party can perform any operation allowed by the laws of
quantum physics, all processes ideally accomplished with no
imperfection of any kind.  During these exchanges, Adam would have
committed a bit with associated evidence to Babe.  It is argued that,
at the end of the commitment phase, there is an entangled pure state
$\ket{\Phi_\sb}$, $\sb \in \{0,1\}$, shared between Adam who
possesses state space $\cH^A$, and Babe who possesses $\cH^B$.  For
example, if Adam sends Babe one of $M$ possible states $\{
\ket{\phi_{\sb i}} \}$ for bit \sb\ with probability $p_{\sb i}$, then
\begin{equation}
\ket{\Phi_{\sb }} = \sum_i \sqrt{p_{\sb i}}\ket{e_i}\ket{\phi_{\sb i}}
\label{eq:entstate}
\end{equation}
with orthonormal $\ket{e_i} \in \cH^A$ and given $\ket{\phi_{\sb i}}
\in \cH^B$.  Adam would open by making a measurement on $\cH^A$, say
$\{ \ket{e_i} \}$, communicating to Babe his result $i_0$ and $\sb$;
then Babe would verify by measuring
the corresponding projector $\ketbra{\phi_{\sb i_0}}{\phi_{\sb i_0}}$ on $\cH^B$,
accepting as correct only the result 1.

More generally, when classical random numbers known only to one party
are used in the commitment, they are to be replaced by corresponding
quantum entanglement purification.  The commitment of $\ket{\phi_{\sb
    i}}$ with probability $p_{\sb i}$ in (\ref{eq:entstate}) is, in
fact, an example of such purification. An example involving Babe is an anonymous state
protocol \cite{yue2}-\cite{yue4} where $\ket{\phi_{\sb i}}$ in
(\ref{eq:entstate}) is to be obtained by Adam applying unitary
operations $U_{\sb i}$ on state $\ket{\psi_k} \in \cH^{B_1}$ sent to him
by Babe with probability $\lambda_k$, $k \in \{1,\ldots,K\}$.
Generally, for any random $k$ used by Babe, it is argued that from the
doctrine of the ``Church of the Larger Hilbert Space'' \cite{gl}, it is to be
replaced by the purification $\ket{\Psi}$ in $\cH^{B_1} \otimes \cH^{B_2}$,
\begin{equation}
\ket{\Psi} = \sum_k \sqrt{\lambda_k} \ket{\psi_k}\ket{f_k},
\label{eq:purif}
\end{equation}
where the $\ket{f_k}'s$ are complete orthonormal in $\cH^{B_2}$ kept
by Babe while $\cH^{B_1}$ would be sent to Adam.  With such purification, it is claimed that any protocol
involving classical secret parameters would become quantum-mechanically
determinate, i.e., the shared state $\ket{\Phi_\sb}$ at the end of
commitment is completely known to both parties.  Note that, from
(\ref{eq:purif}), this means that both $\{\lambda_k\}$ and
$\{\ket{f_k}\}$ are taken to be known exactly to both Babe {\it and Adam}.

Why should Adam and Babe share a pure state instead of a mixed one at
the end of commitment? One key ingredient of the impossibility proof
is the use of measurement purification, or quantum computers, in lieu
of actually taking macroscopic measurement readings.  During
commitment, quantum registers holding the measurement results would be
passed along instead.  Furthermore, any measurement followed by a unitary operation $U_l$ depending on the
measurement result $l$ would be equivalently described by an overall
unitary operator.  Thus, if the orthonormal $\{ \ket{g_l} \}$ on
$\cH^{C_1}$ is measured with result $l$, and then $U_l$ operates on $\cH^{C_2}$, it is
equivalent to the unitary operation
\begin{equation}
U = \sum_l \ketbra{g_l}{g_l} \otimes U_l
\label{eq:unitop}
\end{equation}
on $\cH^{C_1} \otimes \cH^{C_2}$.  It is claimed that any actual
measurement during commitment can be postponed until the opening and the
verification phases of the protocol without affecting the protocol in
any essential way.  In order to maintain quantum
determinacy, the exact $\{ \ket{g_l} \}$ in
(\ref{eq:unitop}) are taken to be known to both parties, even though
the measurement may be chosen by a party among different possible
alternatives.  Let us use
$k$ to denote Babe's secret parameter, and $i$ to denote Adam's secret
parameter, such as the $i$ with probabilities $\{p_i\}$ in
(\ref{eq:entstate}). These
crucial {\it assumptions} of openly known $\{p_i\}$, $\{ \lambda_k \}$, $\{
\ket{f_k} \}$, and $\{ \ket{g_l} \}$ are made in the
impossibility proof through the use of known fixed quantum computers
or quantum machines for data storage and processing by either party
\cite{may2}, \cite{lc3}, \cite[Appendix]{bcml2}, even though the control of such machines belongs
only to one of the parties.

Generally, Babe can try to identify the bit from $\rho^B_\sb$, the
marginal state of $\ket{\Phi_\sb}$ on $\cH^B$, by performing an
optimal quantum measurement that yields the optimal cheating
probability $\bar{P}^B_c$ for her.  Adam cheats by committing
$\ket{\Phi_0}$ and making a measurement on $\cH^A$ to open $i_0$ and
$\sb=1$.  His probability of successful cheating is computed through
$\ket{\Phi_\sb}$, his particular measurement, and Babe's verifying
measurement; the one optimized over all of his possible actions will
be denoted $\bar{P}^A_c$.  For a fixed measurement basis, Adam's
cheating can be described by a unitary operator $U^A$ on $\cH^A$.
Thus, his general EPR attack goes as follows.  For a general protocol, the
shared state $\ket{\Phi_\sb}$ at the end of commitment is not
necessarily of the form (\ref{eq:entstate}), but is nevertheless an
openly known pure state on $\cH^A \otimes \cH^B$.  If the protocol is
perfectly concealing, i.e, $\bar{P}^B_c = 1/2$, then $\rho^B_0 = \rho^B_1$.  By writing
$\ket{\Phi_\sb}$ as the Schmidt decomposition on $\cH^A \otimes
\cH^B$,
\begin{equation}
\ket{\Phi_\sb} = \sum_j \sqrt{\tilde{p}_j} \ket{\tilde{e}_{\sb j}}
\ket{\tilde{\phi}_j},
\label{eq:schmidt}
\end{equation}
where $\ket{\tilde{\phi}_j}$ are the eigenvectors of $\rho^B_\sb$ and
$\{ \ket{\tilde{e}_{\sb j}}\}$ for each $\sb$ are complete orthonormal in
$\cH^A$, it follows that Adam can obtain $\ket{\Phi_1}$ from
$\ket{\Phi_0}$ by a local cheating transformation $U^A$ that brings
$\{ \ket{e_{0j}} \}$ to $\{ \ket{e_{1j}}\}$.
Whatever operations he needs to perform to open, which may involve
identifying his previous operation rather than a state on $\cH^B$, can
be carried out accordingly after the cheating transformation.  Thus
his optimum cheating probability is $\bar{P}^A_c = 1$ in this case.

For unconditional, rather than perfect, security, one demands that
both cheating probabilities $\bar{P}^B_c - 1/2$ and $\bar{P}^A_c$ can
be made arbitarily small when a security parameter $n$ is increased
\cite{may2}.  Thus, {\it unconditional security} is quantitatively expressed
as
\begin{equation}
({\rm US}) \qquad \lim_n \bar{P}^B_c = \frac{1}{2},\quad \lim_n
  \bar{P}^A_c = 0.
\label{eq:us}
\end{equation}
The condition (\ref{eq:us}) says that, for any $\epsilon > 0$, there
exists an $n_0$ such that for all $n > n_0$, $\bar{P}^B_c - 1/2 <
\epsilon$ and $\bar{P}^A_c < \epsilon$, to which we may refer as
$\epsilon$-{\it concealing} and $\epsilon$-{\it binding}.  These
cheating probabilities are to be computed purely on the basis of
logical and physical laws, and thus would survive any change in
technology, including an increase in computational power.  In general,
one can write down explicitly
\begin{equation}
\bar{P}^B_c = \frac{1}{4}\left(2 + \trnorm{\rho^B_0 -
  \rho^B_1}\right),
\label{eq:barpc}
\end{equation}
where $\trnorm{\cdot}$ is the trace norm, $\trnorm{\tau} \equiv \tr
(\tau^\dag \tau)^{1/2}$ for a trace-class operator $\tau$, but the
corresponding $\bar{P}^A_c$ is more involved.  Nevertheless, the
impossibility proof shows that Adam can find a cheating $U^A$ that yields
\begin{equation}
({\rm IP}) \qquad \lim_n \bar{P}^B_c = \frac{1}{2} \,\, \Rightarrow
  \,\, \lim_n \bar{P}^A_c = 1
\label{eq:ip}
\end{equation}
within its formulation \cite{may1,yue2}.  Note that the impossibility
proof makes a stronger statement (IP) than the mere
impossibility of (US), i.e., (\ref{eq:ip}) is stronger than
(\ref{eq:us}) not being possible.

There are various gaps and implicit assumptions hidden in the
impossibility proof, many of which seem to spring from the idea that a
protocol leads to a closed quantum system all by itself, requiring no
interaction with external agents or preparers. These gaps render the proof
incomplete in several ways.  As to be discussed in the following, some
of them can be partially justified or closed, but many still remain
and cannot be bridged.
We will refer to protocols that fit this impossibility proof
formulation as Type 0 protocols, and will describe four additional
types, 1, 2, 3, and 4, that are clearly not covered by this proof.
Before proceeding, we first elaborate on the limited scope of the
impossibility proof formulation.

\newpage

\section{Problems of the impossibility proof}
\label{sec:probs}

\renewcommand{\thefootnote}{\fnsymbol{footnote}}

A plausible first reaction to the impossibility proof is: why are all
possible QBC protocols reducible to the formulation described in the
last section?  More precisely, how may one characterize quantitatively
the necessary feature of an unconditionally secure QBC protocol in
order to show it to be impossible?  To put this in yet another way,
what is the mathematical definition of a QBC protocol, or the
mathematical statement of the necesary feature of an unconditionally
secure QBC protocol, that is required for any proof of a mathematical
theorem that says such protocol is impossible?  {\it No} such definition is
available.  The situation is similar to the lack of a definition of an
``effectively computable'' function.  Since nobody calls the
Church-Turing thesis the Church-Turing theorem, at best the
impossibility proof is a ``thesis'' which may be found incorrect in
the future.  This {\em a priori} logical point is further elaborated
in Appendix A.

The crucial starting point of the impossibility proof asserts that, in
general, a protocol is equivalent to one with openly known pure states
$\ket{\Phi_\sb}$ on $\cH^A \otimes \cH^B$ at the end of commitment.
Let us explore what this entails.  Suppose Adam commits, in a
prescribed protocol, one of $M$ possible $\ket{\phi_{\sb i}}$ for each
$\sb$ without entanglement.\footnote[2]{In practice, this is what
would happen currently due to the difficulties of generating and
maintaining entanglement.  Some of these difficulties are not merely
technical, but are actually inherent in principle.}  Then $\rho^B_0$
is identical or close to $\rho^B_1$ as before, but Adam cannot cheat.
This situation is not one where a pure $\ket{\Phi_\sb}$ is known to
Babe, which occurs only when {\it all} the randomness on $\rho^B_0$
comes from quantum entanglement.  Even then Adam can cheat only if the
entanglement is controlled by him.  Indeed, quantum entanglement is not
a conceptual resource, but rather a physical one, and needs to be
physically established. See Appendix B for a discussion
on randomness generated by quantum entanglement versus that generated
by other means, and the
confusion surrounding the doctrine of ``Church of the Larger Hilbert
Space.''  There may exist protocols in which Adam is forced
to generate randomness without being able to entangle over it during
the course of commitment, so that at the end of commitment one has the
situation described above instead of openly known $\ket{\Phi_\sb}$ on
$\cH^A \otimes \cH^B$.  In our Type 4 protocols, the protocol design
technique of open questioning of evidence could be used to achieve this situation,
as described in Section \ref{sec:type4} and Appendix C.

In general, if Babe makes an actual measurement during commitment,
there would not be an openly known $\ket{\Phi_\sb}$ at the end of
commitment.  The impossibility proof claims that such measurement can
be postponed until after commitment with the use of measurement
purification and (\ref{eq:unitop})
in place of an actual measurement.  However, {\em no} proof is given
that both $\bar{P}^A_c$ and $\bar{P}^B_c$ would not be affected.
Furthermore, one has to make sure that it is not the microscopic
states Babe is thus required to store her measurement results that are
being reversed by Adam's cheating. Otherwise, Babe can take
macroscopic readings instead.  In Section \ref{sec:newprot}, we will
show how cheating detection during commitment is {\it not}
incorporated in the impossibility proof formulation with an openly
known $\ket{\Phi_\sb}$.  As a consequence, the situation of actual
measurements during commitment has to be explicitly included in a
general formulation of QBC protocols.

The use of anonymous states (\ref{eq:purif}) alone, where $\{
\lambda_k \}$ may be unknown to Adam, leads to our Type 3 protocols.  As
elaborated in Appendix B, the distinction between an unknown and a
random parameter is crucial in this sutiation, and the assertion that
$\ket{\Phi_\sb}$ is openly known cannot be maintainted.  A theory of
statistical quantum games is required for an analysis of protocols of
this type.

Assuming that $\ket{\Phi_\sb}$ is openly known at the end of
commitment, it is still not proved that Adam can cheat in general
because special structure or mingling of $\cH^A$ and $\cH^B$ during
commitment may lead to an opening and verification procedure different
from Adam and Babe acting on $\cH^A$ and $\cH^B$ separately. Our
Type 2 protocols give one such possibility, but no doubt there are others. Generally in a QBC protocol with a given
$\ket{\Phi_\sb}$ at the end of commitment, different opening
and verification strategies are possible, depending on exactly how
$\ket{\Phi_\sb}$ is arrived at. Both our Type 2 and Type 4 protocols
may be viewed as ones where these phases are more complex than the one
given in the formulation of the impossibility proof. In the next section we will first elaborate on the issue of what may
constitute a QBC protocol and whether we can give it a mathematical
definition.

\newpage

\section{Proper framework for protocol
  formulation}\label{sec:proper}

The following two principles, the Intent Principle and the Libertarian
Principle, govern the viability and meaningfulness of any bit
commitment protocol in a descriptive, not normative, sense.  That
is, they would be satisfied in what we would take
intuitively to be a proper protocol, and are not imposed in a
legislative fashion, as discussed in the following.

\vskip 0.15in
\begin{quotation}
\noindent{INTENT PRINCIPLE --- Each party would act to achieve the intent of the
protocol if no cheating by the other party is (probabilistically)
possible.}
\end{quotation}
\vskip 0.1in

Thus, each party would cooperate so that the protocol would not be
{\it aborted}, which happens when one party is found cheating by the
other through a possible cheat-detection mechanism during the
commitment phase.  Since each party can always just abort by
noncooperation during any stage of any two-party protocol, the Intent
Principle does not exclude any action not otherwise possible.  Thus,
if the cheating detection probability leads to an overall cheating
success probability within the given $\epsilon$, the protocol is a
proper one and cannot be
declared illegitimate because one party may keep cheating, though
keep being detected.

We also have the

\vskip 0.15in
\begin{quotation}
\noindent{LIBERTARIAN PRINCIPLE --- At any stage of the protocol, each party can freely perform any
  possible local operation consistent with the Intent Principle for cooperation.}
\end{quotation}
\vskip 0.1in

Thus, no party can be assumed to be honest in anything if the action
leads to his/her own advantage and would not get caught.  That is,
each party can cheat whenever possible, unless it violates the Intent
Principle for cooperation.  
There would be {\it no need} for any protocol if the parties can be
assumed honest.  Similarly, each party can do whatever is possible to
thwart the other party's cheating. Under the Intent Principle, a party
is obliged to accept a protocol if he is assured that the probability of
cheating against him is within the tolerance level $\epsilon$, even though he does not know a secret parameter of the other
party.  The following Secrecy Principle is a corollary of the above
two principles.

\vskip 0.15in
\begin{quotation}
\noindent{Corollary (SECRECY PRINCIPLE) --- A party does not need to reveal a
secret parameter chosen by her in whatever manner, if it does not
affect the other party's security.}
\end{quotation}
\vskip 0.1in

\noindent{On the other hand, if a party has no control or checking on
a secret parameter that the other party may use to cheat successfuly,
she would not accept the protocol.

Any finite sequence of two-way quantum communication exchanges that results in
bit commitment under the Intent Principle is evidently a QBC protocol,
whose security is to be analyzed under the Libertarian Principle.
More importantly, any QBC formulation that fails to include all such
sequences does not capture all possible QBC protocols. The present
framework is more general than the ``Yao model'' \cite{yao,lc3} in
that aborting the protocol on the basis of cheating detection is allowed during commitment, and is more
specific in the explicit formulation of the above principles. As
discussed in the preceding section, the impossibility proof formulation is not complete in that it
misses protocols with cheating detection during commitment because
such detection would involve actual measurements that may not be
postponed until after commitment to yield openly known
$\ket{\Phi_\sb}$.  Also, the Secrecy Principle a priori {\em contradicts} directly
the claim of openly known $\ket{\Phi_\sb}$.

The above principles do not constitute a mathematical definition of a
QBC protocol.  They are too broad and too narrow at the same time ---
too broad in the sense that bit commitment is not defined, and too
narrow in that other possibilities may still exist.  My personal
suspicion is that the ``too broad'' problem, or the difficulty of
defining bit commitment, is much more serious than the ``too narrow''
problem.

As in all QBC formulations
so far, it is assumed in this paper that Adam opens perfectly on one
bit value, say $\sb = 0$.  More generally, one may allow QBC protocols
that open on one bit with a success probability $P_0 = 1-\epsilon'$
for a small $\epsilon'$.  It appears that protocols for which neither
bit can be opened with near-unity probability are of little
interest.  In conjunction with $\epsilon$-concealing and
$\epsilon$-binding, one may then consider the possibility of
$(\epsilon,\epsilon')$-protocols, the detailed treatment of which will
be given elsewhere.

\newpage

\section{New protocol techniques, or gaps in the impossiblity proof}\label{sec:newprot}

In this section we describe three new techniques for constructing QBC
protocols, which are not covered by the impossibility proof formulation.
Our Type 1 protocol is based on the first technique, Type 2 on the
second and, possibly, additional others, Type 3 on
random numbers, and Type 4 on the third technique. Each of these protocol types will be discussed
separately in the following sections.

The {\it first} technique introduces testing on states of an ensemble, in
space or in time, submitted by the other party, in order to check
whether only admissible states of the protocol are being used. This
was already utilized in {\bf QBC2} of Ref.~\cite{yue2}.  The
protocol is aborted if cheating is detected by a measurement.  Such
protocols are allowed under the Intent Principle, but {\it not} included in
the impossibility proof formulation for the following reason.  Babe
can use many different possible $U_l$ in (\ref{eq:unitop}), secretly
chosen to be recognized only by her, in order to represent her choice
of aborting the protocol.  Thus, the resulting $\ket{\Phi_\sb}$ is
{\em not}
known to Adam.  Even if the measurement checking is postponed until
verification, there is no proof that the cases of Adam's successful
cheating do not correspond to the ones aborted by Babe.  That is, a
careful analysis of the overlaps between aborting probabilities by
Adam and Babe with $\bar{P}^A_c$ and $\bar{P}^B_c$ is required.  One
also has to rule out the situation where one keeps aborting if he
finds the situation not conducive to his cheating.  Generally, in
accordance with the Intent Principle, a
fixed number $N_c$ of cheating detections may be built into the
protocol, beyond which the whole attempt at a protocol is aborted.  An
appropriate theory of statistical quantum games needs to be developed
for general analysis of such protocols.

For the {\it second} technique, consider a protocol in which Babe forms
(\ref{eq:purif}) and sends Adam $\cH^{B_1}$, with $\ket{\psi_k} =
\ket{\psi_{k1}}\ket{\psi_{k2}}$ in $\cH^{B_1} =
\cH^{B_{11}}\otimes \cH^{B_{12}}$.  Adam randomly
switches the state in $\cH^{B_{11}}$ to be that of $\ket{\psi_{k1}}$ or
$\ket{\psi_{k2}}$  by the unitary perumation $P_m$,
$m \in \{1,2\}$, modulates the resulting state in $\cH^{B_{11}}$ by
a single $U_\sb$ for each $\sb$, and sends it to Babe.  He opens by
revealing $\sb$, his random permuation $P_m$, and {\it returning}
$\cH^{B_{12}}$.  Babe verifies by testing the
apropriate states in $\cH^{B_{11}}$ for checking $\sb$, and
$\cH^{B_{12}}$ for checking that there is no change.  Thus, Adam
cannot entangle and use $\cH^{B_{12}}$.  It is possible that the
protocol is both concealing and binding
because, for the final commitment state $\ket{\Phi_\sb}$ with Adam
entangling the $P_m$ with $\ket{e_i} \in \cH^{A_1}$, we have $\cH^A =
\cH^{A_1} \otimes \cH^{B_{12}}$ and $\cH^B =
\cH^{B_{11}} \otimes \cH^{B_{12}}$.  Thus, $\rho^B_0$ can be close to
$\rho^B_1$ because $\cH^{B_{12}} \otimes \cH^{B_{13}}$ is not
available to Babe for her cheating.  However, only $\cH^{A_1}$, and
not $\cH^A$, is avaiable to Adam's cheating, so he cannot apply the
required cheating $U^A$ without being found cheating with a
nonvanishing probability.  There is no impossibility proof covering
this situation.

\vskip 0.15in
\noindent{{\bf Example 1} (protocol {\bf QBCp2})}
\vskip 0.15in
\noindent{As a specific example, consider the case $\cH^{B_1} = \cH^{B_{11}}
\otimes \cH^{B_{12}} \otimes \cH^{B_{13}} \otimes \cH^{B_{14}}$ of
four qubits, with $\{ \ket{\psi_k} \} =
\{\ket{1}\ket{2}\ket{3}\ket{4}, \ket{4}\ket{1}\ket{2}\ket{3},
\ket{3}\ket{4}\ket{1}\ket{2}, \ket{2}\ket{3}\ket{4}\ket{1}\}$, where
$\{ \ket{1},\ket{2},\ket{3},\ket{4}\}$ are, e.g., a fixed set $S_0$ of four
possible {\bf BB84} states on a given great circle of a qubit.  Adam
permutes each $\ket{\psi_k}$ by one of four possible $P_m$, and returns
the first qubit to Babe unchanged for $\sb=0$, while shifted by $\pi/2$
in the great circie for $\sb=1$.  Assume first that Babe either did
not entangle, or cannot use
her entanglement in $\cH^{B_2}$, so that Adam receives one of the
four possible $\ket{\psi_k}$.  It is then easy to see that
$\rho^{B_{11}}_0(\psi_k) = \rho^{B_{11}}_1(\psi_k)$ for all $k$.  It is also
not hard to see that no entanglement of the four possible $P_m$ would
produce a rotation on the first qubit while not disturbing the
others.  Thus, Adam cannot cheat perfectly and has a fixed
$\bar{P}^A_c$ for this protocol which is not arbitrarily close to one,
even though it is perfectly concealing.  In Section
\ref{sec:space}, we will indicate how Babe can be effectively denied
her use of entanglement via $\cH^{B_2}$.
\vskip 0.15in

In the {\em third} technique, Babe asks Adam some of infinitely many
possible questions concerning the evidence that Adam committed,
demanding the answers to be presented to her in a random fashion, specified
by her as a quantum code.  Since Adam cannot entangle this new random
code on top of the entanglement he already formed, he could only cheat
successfully if the required presentations of the answers have been
pre-entangled by him. However, he can
pre-entangle answers to only a finite number of questions, and thus
can only cheat with an arbitrarily small probability.

Each of these techniques will now be elaborated upon in the different
types of protocols.

\newpage

\section{Some measurements cannot be postponed:  Type 1 protocols}\label{sec:nopost}

In this Section we will show that the protocol technique of testing
for cheating detection alone, with resulting protocols referred to as
Type 1, could already lead to unconditional security.  The general
idea leading to our protocol {\bf QBC1} would be first described
before security analysis and a precise statement of the protocol.

If carried out honestly, the protocol would work as follows.  Adam
sends Babe a large number $n$ of qubits named by their temporal
position with states selected randomly and independently from the set
$S_0$ of four {\bf BB84} states on a given fixed great circle $C$ of
the qubits.  Babe randomly selects $n - n_0$ qubits, tests them by
asking Adam what these states are, and verifies them, with $n$ and
$n_0$ large so that the remaining $n_0$ states would also be
distributed nearly uniformly on $S_0$.  She then picks randomly one of the
remaining $n_0$ states and sends it Back to Adam who would modulate it by $U_0 =
I$ or $U_1 = R(\pi)$, rotation by $\pi$ radians on the circle $C$,
depending on $\sb = 0$ or 1.  He opens by revealing $\sb$ and all the
$n_0$ qubit states.  Babe verifies by checking all the qubits in her possession.  This protocol {\bf
QBC1} is $\epsilon$-concealing and $\epsilon$-binding for the
following reasons.

Adam may entangle each individual qubit he sends in the form
(\ref{eq:purif}) with $\ket{\psi_k} \in S_0$, and then measure $\{
\ket{f_k}\}$ when asked to reveal by Babe.  If he sends in other qubit
states, the chance $\epsilon^A_1$ he would escape detection is
arbitrarily small for large $n-n_0$. If he entangles across qubits, that merely reduces his
freedom in response to Babe's testing.  When he accepts the qubit sent
back by Babe, he would have to measure $\{ \ket{f_k}\}$ in all the
remaining qubits before his modulation, or else he could not commit because there would be
no difference between his two $U_\sb$ actions.  If he
measures on the qubit sent back by Babe, he would not be able to open
perfectly for $\sb=0$. More significantly, the information is of little use to
him since he does not know the name of that qubit. He can only cheat
by declaring $\sb=1$ and switching the names of some of the qubits,
hoping that it would fit his cheating $\sb=1$ opening.  However, the chance
that would succeed without being detected can be seen to be arbitrarily small for
large $n_0$.

This protocol is $\epsilon$-concealing because all of Babe's possible
cheatings would be unsuccessful as follows. With a high probability,
Adam checks the qubit sent back by Babe with a question on its name,
and verifies it is correct.  He would accept the qubit at some point.
If Babe sends in a state
different from one in Adam's ensemble, the probability that would
not get detected is arbitrarily small when Adam tests a large number $m \ll
n_0$ of times. Assuming that both Adam and Babe employ a randomized
strategy applied independently from qubit to qubit during Adam's
testing, it can be readily shown that the protocol is
$\epsilon$-concealing and $\epsilon$-binding for sufficiently large
$n$ and $n_0$. This
different-state attack by Babe includes her possible entanglement,
even though it can be shown independently that her entanglement would
not help.  She can also try to determine the qubit state she sends
back by measuring the other qubits in her possession, but these are not
correlated to the qubit she sends back. We have the folowing protocol
{\bf QBC1}.

\begin{center}
\vskip 0.15in
\framebox{
\begin{minipage}{5in}
\vskip 0.1in
\underline{PROTOCOL {\bf QBC1}}

{\small \begin{enumerate}
\item[(i)] Adam sends Babe a large number $n$ of independent qubit
states drawn randomly from $S_0$, a set of openly known {\bf BB84} states on a
given great circle $C$ of the qubits. The qubits are named by their
temporal positions as received by Babe.
\item[(ii)] Babe randomly picks a large number $n_0$ of these qubits,
sets them aside, and asks Adam to open the remaining ones.  She
verifies them to be correct and distributed nearly uniformly, as
prescribed in (i).  Otherwise the protocol is aborted.
\item[(iii)] Babe sends back one of the $n_0$ remaining qubits to
Adam, who checks it a sufficient number of times in a game with Babe,
accepts one, and modulates it by either $U_0 = I$ or $U_1 = R(\pi)$, rotation
by $\pi$ radians on $C$, and sends it back to Babe.
\item[(iv)] Adam opens by revealing $\sb$ and all the remaining qubit
states.  Babe verifies by measuring the corresponding projectors.
\end{enumerate}
\vskip 0.1in
}
\end{minipage}
}
\end{center}
\vskip 0.2in

In addition to being an outline, the above security analysis is
incomplete because the optimal sequential decision in both Adam's and
Babe's testing have not been analyzed.  A new development of quantum
sequential decision theory and quantum games is needed for such an analysis.  In
general, a fixed number $N_c$ of cheatings by each party is allowed as
a protocol design parameter in a quantum game situation.  A party is
not permitted, i.e., loses the game, if found cheating more than $N_c$
times.  It is evident that {\bf QBC1} is unconditionally secure if
$N_c$ is taken to be zero or a small number. I believe that, for any
given $N_c$, $n$ and $n_0$ can be chosen so that the protocol is
unconditionally secure for any $\epsilon > 0$.

Note that {\bf QBC1} is not just a cheat-sensitive \cite{hk}
protocol.  In particular, the cheat detection is done before the bit
is committed.  As shown in the preceding section, it would not be
equivalent to a protocol with an openly known $\ket{\Phi_\sb}$ at the
end of commitment.

\newpage

\section{Who has which space:  Type 2
protocols}\label{sec:space}

The use of the first technique in Section \ref{sec:newprot}, test for
cheating via measurement, has the effect of changing and pinning down
the $\epsilon$-concealing condition of the protocol, as compared to
one without the test.  Generally, the condition
\begin{equation}
\rho^B_0(\Psi) \stackrel{\epsilon}{\sim} \rho^B_1(\Psi)\qquad {\rm
for\,one\,} \ket{\Psi} \in \cH^{B_1} \otimes \cH^{B_2},
\label{eq:con1}
\end{equation}
while weaker than
\begin{equation}
\rho^B_0(\Psi) \stackrel{\epsilon}{\sim} \rho^B_1(\Psi)\qquad {\rm
for\,every\,} \ket{\Psi} \in \cH^{B_1} \otimes \cH^{B_2},
\label{eq:con2}
\end{equation}
is not equivalent to
\begin{equation}
\rho^B_0(\psi_k) \stackrel{\epsilon}{\sim} \rho^B_1(\psi_k)\qquad
\forall\,\ket{\psi_k} \in \cH^{B_1}.
\label{eq:con3}
\end{equation}
Specifically, (\ref{eq:con1}) does not imply (\ref{eq:con3}) because
there can be a $\ket{\psi_1}$ for which $\rho^B_0(\psi_1)$ and
$\rho^B_1(\psi_1)$ are far apart under (\ref{eq:con1}) with
$\lambda_1$ small \cite{yue4}.  Also, it is easy to check that, in
Example 1 of Section \ref{sec:newprot}, (\ref{eq:con3}) holds with
equality, but there is a finite gap for $\trnorm{\rho^B_0 - \rho^B_1}$
upon entanglement with $\cH^{B_2}$.  This renders {\it false} the claim that
the use of random numbers as in (\ref{eq:con3}) can be equivalently described by their quantum
purifications as in (\ref{eq:con1}).  Further discussion os this point is given in Section
\ref{sec:type4}.  Here we note that (\ref{eq:con2}) is, in general,
a sufficient but not necessary (at least not having been proved
necessary) condition for the protocol to be concealing, again to be
further discussed in Section \ref{sec:type4}.  It is rather a severe
restriction on the protocol that can be relaxed to (\ref{eq:con1})
with test for cheating.

A Type 2 protocol involving also the first technique of cheating
detection may work as follows.  Similar to {\bf QBC1}, a large $n$-sequence
($n$-fold tensor product) of qubit states, drawn independently with
probability $\lambda_k$ from a fixed set $S_0 = \{ \ket{\psi_k}\}$,
would be sent from Babe to Adam, each state named by its position in
the sequence.  Adam puts aside randomly chosen $n_0$ of them, and asks
Babe to reveal the remaining $n-n_0$ ones for testing.  For large
enough $n-n_0$, Babe cannot use any $\ket{\Psi} \in \cH^{B_1} \otimes
\cH^{B_2}$ other than that of the form (\ref{eq:purif}) without
getting caught with probability arbitrarily close to one, so that the
concealing condition is (\ref{eq:con1}), and not (\ref{eq:con2}).
If Adam randomly picks one of the remainng $n_0$, or $m$ for full
unconditional security, modulates it by a
single $U_\sb$ for each $\sb$, and return it without the name to Babe,
she would not be able to use her entanglement (\ref{eq:con1})
effectively on any qubit.  This technique is similar to the use of
decoy states from Adam to Babe in \cite{yue4}, and results in an
effective concealing condition (\ref{eq:con3}) in place of
(\ref{eq:con1}), although (\ref{eq:con1}) still applies overall. While the
use of a single $U_\sb$ does not allow Adam to cheat successfully on a
fixed qubit, the freedom from the $n_0$-ensemble still allows him to
entangle and launch an EPR attack.  This attack is thwarted via the
second technique of Section \ref{sec:newprot}, which demands that Adam
return the remaining $n_0-1$ qubits so Babe can verify that they have
not been disturbed.  Example 1, our protocol {\bf QBCp2}, can be
extended in this way to become an unconditionally secure protocol {\bf
QBC2}, which is a modified version of a protocol with the same name in
Ref.~\cite{yue2}.  Alternatively, the same logic applies to the following
protocol, which is somewhat simpler.

\begin{center}
\vskip 0.15in
\framebox{
\begin{minipage}{5in}
\vskip 0.1in
\underline{PROTOCOL {\bf QBC2A}}

{\small \begin{enumerate}
\item[(i)] Babe sends Adam $n$ qubits named by their temporal position,
each drawn independently with equal probability from $S_0$, a fixed
set of four possible {\bf BB84} states.
\item[(ii)] Adam randomly picks $n_0$ of these qubits and sets them
aside, and asks Babe to open the remaining $n-n_0$ ones.  He
verifies them to be correct in that they are distributed as prescribed
in step (i).  Otherwise the protocol is aborted.
\item[(iii)]Adam randomly picks $m$ out of the $n_0$
remaining ones, modulates each by the same $U_0 = I$ or $U_1 = R(\pi)$, rotation
by $\pi$ on the great circle containing $S_0$, and sends them back to
Babe.
\item[(iv)] Adam opens by revealing $\sb$ and returning the remaining
$n_0 - m$ qubits.  Babe verifies by measuring the corresponding projectors.
\end{enumerate}
\vskip 0.1in
}
\end{minipage}
}
\end{center}
\vskip 0.2in

By proper choice of $m$, $n_0$, and $n$, this protocol can be made both
$\epsilon$-concealing and $\epsilon$-binding for any $\epsilon > 0$,
given that Adam opens perfectly on $\sb=0$.  The main steps of the
proof may be outlined as follows.  Babe can cheat by entangling over
each individual qubit and also by using a distribution of qubits more biased
than the one presented in step (i).  To defeat her qubit entanglement
cheating, let $n_0/n = \epsilon_1$.  The probability that she would
pair $\cH^{B_{11}}$ with the correct $\cH^{B_{21}}$, where $\cH^B =
\cH^{B_1} \otimes \cH^{B_2}$, $\cH^{B_m} = \cH^{B_{m1}} \otimes \ldots
\otimes \cH^{B_{mn}}$, $m \in \{1,2\}$, is thus $\epsilon_1$.  If the
pairing is incorrect, the trace distance in (\ref{eq:ip}) is not
affected because, for any three general states $\rho,\rho',\sigma$,
\begin{equation}
\trnorm{(\rho - \rho') \otimes \sigma} = \trnorm{\rho - \rho'}.
\label{eq:tnmult}
\end{equation}
If the pairing is correct, we take the upper bound value of two for
the trace distance.  By making both $n_0$ and $n$ large and testing on
the arbitrary $n-n_0$ qubits, one may guarantee, to within any
$\epsilon_2 > 0$ for the resulting $P^B_c = 1/2+\epsilon_2$ with
$\epsilon_2 \rightarrow 0$ in the limit $n_0 \rightarrow \infty$ and $n \rightarrow
\infty$, that the distribution of states in the two sets of qubits is
indeed the one prescribed.  Accordingly, Babe can only get $P^B_c =
1/2 + \epsilon_3$ for $\epsilon_3 \rightarrow 0$ from the $m$
committed qubits for any fixed $m$.  This situation has been analyzed
for {\bf QBC2} in Ref.~\cite{yue2}.  From the union bound on probability, one
may take $\epsilon_1 + \epsilon_2 + \epsilon_3 \le \epsilon$, and the protocol
becomes $\epsilon$-concealing.  The asymptotic situation at $m,n_0,n
\rightarrow \infty$ is quite apparent even in the absence of any
quantification with respect to the $\epsilon$'s.  The
protocol is binding on Adam, because $m$ can be chosen large enough so
that Adam's optimum one-qubit cheating probability $p_A$ becomes
$p^m_A \le \epsilon$.

This {\bf QBC2A} utilizes the first technique and denial of
entanglement matching, in addition to its use of the
second technique, which makes it Type 2.  Even though there is yet no example, one cannot
{\it a priori} rule out the possibility that the use of the second
technique alone, as described in Section \ref{sec:newprot}, would lead
to an unconditionally secure protocol.  Even if that turns out to be
impossible, the impossibility proof formulation does not cover such
situation, and need to be extended for a proof.

\newpage

\section{Classical randomness and quantum
  purification:  Type 3 protocols}\label{sec:random}

A cornerstone of the general impossibility proof is the assertion that
classical randomness can be equivalently described as quantum
determinacy via purification, say by (\ref{eq:purif}), through the
doctrine of ``Church of the Larger Hilbert Space,'' a technique also
widely used in quantum coin tossing.  But equivalent for what?  In the
following, we analyze the ways in which they are not equivalent for
use by Babe in a QBC protocol.  The best argument I know for their
equivalence would be given alongside.  Appendix B is essential for
clarification of this issue.

First of all, it is clearly not true that all classical randomness can
be reduced to that arising from quantum description of a system.
After all, there were many scenarios for the occurrence of classical
randomness before the rise of quantum physics, including especially
classical statistical mechanics.  Even if one grants a determinate
quantum description for the underlying classical randomness involved,
it is unreasonable to assume that any party would possess the detailed
knowledge to write down the complete quantum description.  However, in
the context of QBC protocols, it is not only reasonable, but, in fact,
mandatory to consider such purification (\ref{eq:purif}) for which a party can form and use such purification for
cheating.  Thus it is a consideration of entanglement cheating, not the ``Church
of the Larger Hilbert Space,'' that compels one to consider
(\ref{eq:purif}).

The following argument, in the spirit of the impossibility proof,
appears to show that the exact $\{ \ket{f_k} \}$ in (\ref{eq:purif}) need not be known by Adam for
finding his cheating transformation.  Let the protocol be
$\epsilon$-concealing as a consequence of $\rho^B_0(\Psi)$ being close
to $\rho^B_1(\Psi)$ for one $\ket{\Psi}$ generated by Babe in the
form (\ref{eq:purif}).  Assume Babe verifies by first measuring $\{
\ket{f_k} \}$ and then checking Adam's opening.  The commutativity of
Adam's and Babe's operations shows that the protocol performance is
the same whether Babe measures $\{ \ket{f_k} \}$ during commitment or
after Adam opens.  The fact that Adam can cheat after Babe measures
$\{ \ket{f_k} \}$ shows that the cheating must be independent of the
specific $\{\ket{f_k}\}$, even though it is obtained for a known $\{
\ket{f_k} \}$.  Note that this argument does not extend to the
knowledge of $\{ \lambda_k\}$.

Nevertheless, even just for $\{ \ket{f_k} \}$ this argument contains a
major gap, which is, in fact, a general gap in the impossibility
proof:  it is not guaranteed that there is only one veryfing
measurement for the protocol.  In the particular case of randomness
described above, it means that the split measurement of $\{\ket{f_k}\}$
of $\cH^{B_1}$, and then a measurement on $\cH^{B_2}$, is {\it not}
the verifying measurement of $\Pi$ that has been proved susceptible to
cheating as prescribed by the impossibility proof. It is not true
that whenever $\Pi$ is verified on a cheating state, then so is the
split measurement. The cheating probability $\bar{P}^A_c$ depends
on the verifying measurement. For an arbitrary
protocol, the impossibility proof formulatoin does not, and, in fact,
cannot specify what the possible verifying measurements could be.  There is
{\it no} proof given that there cannot be more than one verifying
measurement, for which different cheating transformations are needed.
It turns out that for several types of protocols, though not for all, I can
prove that this is indeed the case in the sense (IP) of (\ref{eq:ip})
for all {\it perfectly verifying measurements}, i.e., measurements
that yield the ``yes'' result with probability one corresponding to
the opening bit value.

However, condition (\ref{eq:con1}), which is taken to be the
$\epsilon$-concealing condition in the impossibility proof, is {\it not} a
proper concealing condition due to the Libertarian Principle.  Indeed,
while (\ref{eq:con1}) implies that Adam can cheat according to the impossibility
proof, the situation is misrepresented in that it may be Babe who can
actually cheat by using a different $\ket{\Psi}$.  Two examples are given in Ref.~\cite{yue4}. It makes
no sense to insist that Babe has to stick to a prescribed $\{
\lambda_k \}$, in contradiction to the Secrecy Principle, so that the
protocol is concealing and Adam can cheat, while Babe can actually use
a different $\{ \lambda_k \}$ and instead cheat successfully
herself. There is no reason for Babe to commit such {\it bit suicide}.
For any protocol, one cannot simply say that a protocol is now taken
to be $\epsilon$-concealing.  One has to describe quantitatively a necessary
$\epsilon$-concealing condition for the protocol before any meaningful
performance analysis can be made, which is something the impossibility proof fails
to do in general. Thus, there is {\em no} impossibility proof whenever
anonymous states $\ket{\Psi}$ are used in a protocol.

Suppose that condition (\ref{eq:con2}) is to be used, which is a
sufficient condition that has not been shown to be necessary for
concealing, as to be discussed later.  For a class of anonymous-state protocols that are
perfectly concealing, it may be shown \cite{yue4,ari,oza} that the
cheating $U^A$ is independent of any $\{\lambda_k\}$ and
$\{\ket{f_k}\}$ in (\ref{eq:purif}). The reason why secure protocols
based on classical random numbers alone are hard to construct is not
necessarily because one forgets quantum purification.  It is because concealing
under quantum purification is often more restrictive than concealing
under classical randomness, as in protocol {\bf QBCp2} or
Example 1. We say that a protocol is of Type 3 whenever states of the
form (\ref{eq:purif}) are used by Babe.  An example is {\bf QBC3} of
Ref.~\cite{yue4}.  Under $\epsilon$-concealing (\ref{eq:con2}) for
such protocols, it is not known whether $\bar{P}^A_c$
is close to one independently of $\{\lambda_k\}$.  Thus, there is {\it
no} impossibility proof if (\ref{eq:con1}) in a Type 0 protocol with
(\ref{eq:purif}) is replaced by (\ref{eq:con2}) or just (\ref{eq:con3}). For such Type 3 protocols, unconditional security may
arise in the following way.  Since Adam does not know $\{\lambda_k\}$,
one may consider first a fixed $\{\lambda_k\}$ and then average over all possible cheating
$U^A$.  Such an average cannot produce $P^A_c \sim 1$. The performance analysis
for the overall situation seems rather involved, and new approaches
may be needed to see whether security is actually provable.  A direct
approach to the analysis of such protocols is given in
Ref.~\cite{ari}.  However, what we have here is actually a
game-theoretic situation involving freedom on both sides with opposing
objectives with regard to the performance criteria $P^A_c$ and
$P^B_c$. It is most appropriate to regard $\{\lambda_k\}$,
$\{p_i\}$, etc.  as unknown with no meaningful
distribution on them, a situation that happens in many problems of
classical statistics whenever there is a lack of statistical regularity or
meaningful ensemble, the situation we have here.  See Appendix B for
further elaboration.

It is argued in \cite{may3} that $\{\lambda_k\}$ has to be taken
openly known in a meaningful protocol, because there is no guarantee
that it can be kept secret. In any cryptographic protocol, one has to assume that anything one
party does on her locality is not known to another party in a distant
locality, relativity or not, or else nothing can be a secret,
including a secret key.  The issue is not why Adam does not
know $\{\lambda_k\}$.  It is why he would know.  Indeed, one may use
the same reasoning and assume Babe knows $U^A$ so she can defeat
Adam's cheating.  The actual situation is that $\{ \lambda_k \}$ is an
unknown parameter in an infinite-dimensional space over ${\mathbb R}$
or ${\mathbb C}$, as discussed in Appendix B.  The conclusion arrived at above can also be repeated in this regard.  Under proper
concealing (\ref{eq:con2}), there is no need for Adam to know $\{
\lambda_k \}$ in accordance with the Secrecy Principle.  There is no
way Adam can find out which particular $\{\lambda_k\}$ Babe uses.  It
is entirely her private affair.  Without (\ref{eq:con2}), Babe is not
going to commit bit suicide with (\ref{eq:con1}).  She would use a different
$\{\lambda_k\}$ instead.

Generally, it is difficult to pin down a necessary condition
for $\epsilon$-concealing for an arbitrary protocol without utilizing
specific information about the protocol details. In fact, the very
meaning of concealing in an arbitrary protocol has to be decided upon. Thus,
(\ref{eq:con2}) may be too strong because Babe in general does not
know the distribution $\{p_i\}$ on Adam's secret parameter $i$.  It
may not be necessary for $\epsilon$-concealing that $\rho^B_0
\stackrel{\epsilon}{\sim} \rho^B_1$ holds for {\it any} $\{p_i\}$ due
to averaging or to the game situation involving $\{\lambda_k\}$ just discussed.
Thus, a general impossibility proof for Type 3 protocols would face
the immediate obstacle of not being able to specify quantitatively
either a {\it necessary} $\epsilon$-concealing or $\epsilon$-binding condition.  One
the other hand, security proof for a particular protocol is much
easier because {\it sufficient} conditons and protocol mechanism can be
specifically exploited.

We summarize the main points concerning random numbers.
\begin{enumerate}

\item Classical randomness is not generally reducible to quantum
uncertainty.

\item The condition of $\epsilon$-concealing with random numbers is
not equivalent to its quantum purification version, i.e.,
(\ref{eq:con1}) is not equivalent to (\ref{eq:con3}).

\item The coefficients $\{ \lambda_k \}$ in the quantum purification
(\ref{eq:purif}) are generally not known to the other party.

\item The concealing condition (\ref{eq:con1}) used in the
impossibility proof is, in general, neither necessary nor sufficient
for concealing.

\item There is no general impossibility proof when anonymous states
are involved in a protocol.

\item With random $k$ and $i$, it is difficult to formulate a necessary
$\epsilon$-concealing or $\epsilon$-binding condition in order to start an
impossibility proof.

\item The general situation of an unspecified protocol, even the
simple case (\ref{eq:entstate}), is game-theoretic.
\end{enumerate}

\newpage
\section{Too many possible questions to entangle:  Type 4 protocols}
\label{sec:type4}

In Type 4 protocols, the unentangled state $\ket{\phi_{\sb i}}$ is
brought about from the entangled openly known $\ket{\Phi_\sb}$ through
the asking of questions related to the evidence by Babe.  As a
consequence, one arrives at the situation discussed at the beginning
of Section \ref{sec:probs}, where the randomness that makes up
$\rho^B_\sb$ is not entangled under Adam.  The ideas and procedure are
best explained for the specific case of protocol {\bf QBC4} in the
following.

Adam sends Babe a sequence of $n$ qubits, each in either one of
$\{\ket{\phi},\ket{\phi '}\}$, such that an even number of $\ket{\phi
'}$ corresponds to $\sb=0$, and an odd number to $\sb=1$.  As shown in
Appendix C, the protocol is $\epsilon$-concealing for large $n$ for
any $|\braket{\phi}{\phi '}|^2 = \epsilon_1$, and Adam has the usual
EPR cheat with the entanglement
\begin{equation}
\ket{\Phi_0} = \sum_i \sqrt{p_i} \ket{e_i} \ket{\phi_{0 i}}
\label{eq:eprcheat}
\end{equation}
for $p_i = 1/2^{n-1}$.  It was suggested in v2 of this paper (
quant-ph/0207089v2 ) that Babe now asks Adam to reveal to her $n-n_0$ qubits,
 randomly selected out of $n$, with $n_0$ remaining ones
sufficient to ensure $\epsilon$-concealing. The idea is to force him
to measure the $\{ \ket{e_i}\}$ in (\ref{eq:eprcheat}) to pin down a
specific $\ket{\phi_{0i}}$, thus destroying the entanglement.
However, Adam can respond as follows.  Let $i =
(i_1,\ldots,i_n)$, $i_l \in \{0,1\}$, $l \in \{1,\ldots,n\}$, $\ket{\phi_{0i}} =
\ket{\phi_{0i_1}}\ldots \ket{\phi_{0i_n}}$, $\ket{\phi_{0i_l}} \in \{
\ket{\phi},\ket{\phi '}\}$ in each $\cK_{l2}$, $\cH^B = \bigotimes_l \cK_{l2}$. Then
$\ket{\Phi_0}$ can be extended through local operation to
\begin{equation}
\ket{\Phi'_0} = \sum_i \sqrt{p_i} \ket{i_1} \ldots \ket{i_n} \ket{e_i}
\ket{\phi_{0i}}
\label{eq:eprext}
\end{equation}
in $\cH^{A'} \otimes \cH^A \otimes \cH^B$, $\cH^{A'} =
\bigotimes^n_{l=1} \cH_{l2}$ a product of qubits, and $\braket{i_l =
0}{i_l = 1} = 0$ for each $l$ are two orthogonal states from the {\bf
BB84} state set $S_0$ on a fixed great circle $C$ of each qubit, with
$\ket{i_l = 0}$ corresponding to $\ket{\phi}$ and $\ket{i_l =1}$ to
$\ket{\phi '}$.  In response to Babe's question on a subset $S \subset
\{ i_1,\ldots,i_n \}$, $|S| = n - n_0$, Adam sends Babe the state spaces
$\cH_{l2}$ for all $l \in S$.  Babe can measure on $\bigotimes_{l \in
S}\cH_{l2}$ to find the answer to her question and verify on the
committed $\{ \ket{\phi_{0i_l}} \}$ in $\cK_{l2}$'s.  Since the protocol is
concealing with Babe posessing $\cH^B \otimes \left(\bigotimes_{l \in
S} \cH_{l2}\right)$ and Adam possessing $\cH^{A'} \equiv \cH^A \otimes
\left(\bigotimes_{l \in S^c}\cH_{l2}\right)$, $S^c = \{1,\ldots,n\} -
S$, Adam can cheat
successfully by finding the proper cheating transformation $U^{A'}$ on
$\cH^{A'}$.

Following our same idea, the protocol is now extended as follows. Babe may ask Adam to do the following instead.  He is going
to provide Babe with the $\ket{i_l}$ for all $l$, but with each
$\ket{i_l}$ turned with probability $1/2$ on the great circle $C$ to
the other two orthogonal {\bf BB84} states for $\sb=0$ and
$\sb=1$. That is, with $S_0 = \{\ket{1},\ket{2},\ket{3},\ket{4}\}$
where $\braket{1}{3} = \braket{2}{4} = 0$, the $\ket{i_l}$ is equally
probable to be $\{\ket{1},\ket{2}\}$ for a $\ket{\phi}$, and
$\{\ket{3},\ket{4}\}$ for a $\ket{\phi'}$. The distribution would be
across the $n$ $\ket{i_l}$'s.  It is easy to see, as shown in Appendix C, that the protocol remains
$\epsilon$-concealing for any $\epsilon_1$ by making $n_0$
sufficiently large.  Now Babe asks Adam to reveal a random set of
$n-n_0$ $\ket{i_l}$'s and verifies them on her corresponding
$\ket{\phi_{0i_l}}$'s, making sure that all states in $S_0$ appear
within, say, the Chernov bound limit. Since Adam has not entangled over the above
randomness, from his point of view $\rho^B_0$ and $\rho^B_1$ are not close at all.  Let $F$ be the
Uhlmann fidelity $\tr \sqrt{(\rho^B_0)^{1/2}\rho^B_1(\rho^B_0)^{1/2}}$
between $\rho^B_0$ and $\rho^B_1$ under (\ref{eq:eprcheat}), and $F'$
for the situation just described.  It is difficult to evaluate $F'$,
but one can bound it loosely by $F' \le \frac{1}{\sqrt{2}}F$ from the fact that one has
(\ref{eq:eprext}) now with a single tensor product extension
$\bigotimes_l \ket{i_l}$ for each $\ket{\phi_{0i}}$ constructed from
$S_0$.  The factor $\frac{1}{\sqrt{2}}$ comes from the largest value
of a single overlap
$\braket{i_l = 0}{i_l = 1}$, the minimum between $\ket{\phi_{0i}}$ and
$\ket{\phi_{1j}}$. From (\ref{eq:fid}) in Appendix C, Adam's optimal
cheating probability satisfies $\bar{P}^A_c \le F_A(\rho^B_0,\rho^B_1)$ with
$F_A$ computed from his point of view. Thus, we have arrived at $\bar{P}^A_c
\le \frac{1}{\sqrt{2}}$, contradicting the impossibility proof assertion (IP)
of (\ref{eq:ip}).  By utilizing this scheme repeatedly, one may obtain
an unconditionally secure protocol, as in the way described in {\bf
QBC2} of \cite{yue2}.  However, we can also achieve unconditional
security as follows.

Babe now asks Adam to present $\ket{i_l}$ in two equally probable
states $\{\ket{1},\ket{5}\}$ with $\ket{5} = R(\theta)\ket{1}$ for $\ket{\phi}$ or $i_l = 0$, and in
$\{\ket{3},\ket{6}\}$ with $\ket{6} = R(\theta)\ket{3}$ for
$\ket{\phi'}$ or $i_l =1$. The angle $\theta$ is chosen so small
that the overlap $\braket{5}{3}$ is $\epsilon$ instead of $\frac{1}{\sqrt{2}}$ in the case of
$S_0$.  Thus $\bar{P}^A_c \le \epsilon$, while $\bar{P}^B_c$ can still
be made $\le \epsilon$ by having $n_0$ sufficiently large.  Babe would
ask Adam to reveal $n-n_0$ of these $\ket{i_l}$'s and verify them as above.
Adam's probability of successful cheating by whatever action is exponentially
small in $n-n_0$, but we do not need to quantify it under the
assumption he needs to open perfectly for $\sb=0$.  That would only
occur if he measures $\{ \ket{e_i} \}$ in (\ref{eq:eprext}) to
answer.  Babe can even postpone the measurement to the verification
phase. 

Adam cannot change (\ref{eq:eprext}) to one that allows him to cheat
from pre-entanglement on answers to the questions, due to local state
invariance as follows.  Consider just the pair of qubit states at the
same $l$th position $\ket{i_l}\ket{\phi_{0i_l}}$ with all other qubits
fixed.  When a measurement of $\{\ket{1},\ket{3}\}$ is performed on
$\cH_{l2}$, the state on $\cK_{l2}$ is a superposition of
$\ketbra{\phi}{\phi}$ and $\ketbra{\phi'}{\phi'}$ under
(\ref{eq:eprext}) when averaged over the measurement results.  If the
$\ket{i_l}$ is entangled to $\ket{\phi_{0i_l}}$ over all four states
in $S_0$ or $\{\ket{1},\ket{5},\ket{3},\ket{6}\}$ in the required form
\begin{equation}
\sum^4_{i_l=1}\ket{i_l}\ket{\phi_{0i_l}}
\label{eq:entform}
\end{equation}
with the proper matching $\ket{\phi_{0i_l}}$, such a measurement would
produce a state on $\cK_{l2}$ with non-vanishing
interference terms $\ketbra{\phi}{\phi'}$, $\ketbra{\phi'}{\phi}$ that
can be easily computed.  No local operation on $\cH^{A'}$ can change
the state of $\cK_{l2}$ this way, as a consequence of local state
invariance stated in Appendix C.  Intuitively, it is clear that Adam
cannot entangle on top of entanglement.  Mathematically, he can extend
(\ref{eq:eprext}) as a tensor product, but not as a direct sum.

More generally, Babe can ask Adam to present his answers for each
qubit in any coded form in one qubit, two qubits, or $m$ qubits. She can adjoin an integer
$m$ to precede $i$ in the binary representation with $m+i$ bits,
present Adam with her secretly chosen computable and invertible arithmetical function
$f : \mathbb{N} \rightarrow \{0,1\}$, and ask for the answer
$f(\{m,i\})$.  If Adam is honest and submitted unentangled
$\ket{\phi_{0i}}$, he could answer straightforwardly.  With his
entanglement, he could do the quantum computation in
\begin{equation}
\ket{\Phi'_0} = \sum_i \sqrt{p_i} \ket{f(\{m,i\})}\ket{e_i}\ket{\phi_{0i}},
\label{eq:qcomp}
\end{equation}
passing $\cH^{A'}$ with state $\ket{f(\{m,i\})}$ to Babe.  Babe can
ask for presentations as in the
parity case above.  Indeed, the possibilities of presenting the
individual parity answers in different spaces alone lead to an
infinite set of possible questions of arbitrarily large cardinality.

To be able to cheat under such open questioning, Adam has to
pre-entangle the answer to {\em every} possible question.  However, he
can only pre-entangle the answers to a finite number of questions.
Indeed, even if he pre-entangles an infinite number, he cannot locate
the answer by an algorithm, say the set of Turing-computable functions
alone is already not recursively enumerable.  For the parity function, Babe
can ask questions involving subset parities on the $n$ qubits, already
generating $2^n$ types of questions that is too many to entangle for
$n > 410$ even if one can use all the physical resources in the
Universe.  See Appendix D for this physical limit.

As elaborated in Appendix B and in Sections
\ref{sec:probs}-\ref{sec:newprot}, Babe can automatize a secretly
chosen rule to specify the questions she may ask.  If one wants to
talk about probability, it is fair to say that the probability her
question was pre-entangled by Adam is arbitrarily small.
Alternatively, to avoid unfruitful terminological debate, one may just
say that Adam can cheat if Babe's questions drawn from an infinite set
fall under Adam's finite set of pre-entangled questions.  This
situation was not realized before, and serves effectively to
produce an unconditionally secure protocol.  It may be {\em
emphasized} that if $\ket{\Phi_\sb}$ is openly known, Babe can always
ask a further question that is {\em not} pre-entangled in it, thus
rendering it unentangled, as discussed above.

\begin{center}
\vskip 0.15in
\framebox{
\begin{minipage}{5in}
\vskip 0.1in
\underline{PROTOCOL {\bf QBC4}}

{\small \begin{enumerate}
\item[(i)] Adam sends Babe a sequence of $n$ qubits, each being either one
of $\{ \ket{\phi},\ket{\phi'}\}$, such that an even number of
$\ket{\phi'}$'s corresponds to $\sb=0$, and an odd number to $\sb=1$.
\item[(ii)] Babe asks Adam questions that are sufficient to pin down
the total committed state, requiring him to present his answers in a
specific randomized form.
\item[(iii)] Babe verifies some of the answers by further questions on
how Adam actually randomized them.
\item[(iv)] Adam opens by revealing all the unknown states to Babe.
She verifies by corresponding measurements.
\end{enumerate}
\vskip 0.1in
}
\end{minipage}
}
\end{center}
\vskip 0.2in

This protocol bears a resemblance to a Type 3 protocol, in which Babe
uses a parameter unknown to Adam.  The difference is that an
additional technique, an open questioning of evidence, is used to
guarantee $\epsilon$-concealing for each unknown value that would
require a different cheating arrangement by Adam, something that is
difficult to achieve by means of anonymous states alone.  Also, there
is no discrete approximation to the infinite set of possibilities in
this case, in contrast to the probabilities $\{ \lambda_k\}$.
Furthermore, when Adam misses the value, his cheating probability is
vanishingly small, in constrast to a mistmath between $U^A$ and $\{
\lambda_k\}$.  

To recapitulate the logic of Type 4 protocols: by asking open
questions concerning the evidence with answers presented in a specific
randomized form chosen secretly by her, Babe ensures that Adam can only cheat
successfully by pre-entangling the whole question correctly.  However,
he can do that only with a vanishingly small probability. It is appropriate to emphasize that this type of protocols shows that,
other than unknown parameters, the specifics of a protocol may play a significant role in rendering untenable the assertion that
an openly known $\ket{\Phi_\sb}$ is obtained at the end of
commitment.  This issue has not been adequately addressed in the
impossibility proof.

\newpage
\section{Summary and conclusion}\label{sec:concl}

If there is a general impossibility proof for secure QBC, one should
be able to apply it schematically to any proposed QBC protocol to show
that it is insecure.  This often cannot be done. The reason is that the
impossibility proof formulation is quite restrictive, and many
nontrivial details in a systematic proof have not been spelled out.  Some such criticisms have
already been discussed in Ref.~\cite{yue2}, but they are analyzed
quantitatively in this paper.

We introduced several new
techniques for protocol design, not covered by the impossibility proof
formulation which only applies to what we call Type 0 protocols.  We
presented three new types of protocols:

\begin{itemize}

\item Type 1 --- measurement for cheating detection,
\item Type 2 --- shifting of evidence state spaces,
\item Type 3 --- utilization of anonymous states,
\item Type 4 --- open questioning of evidence.

\end{itemize}

A specific Type 4 protocol, {\bf QBC4}, is proved unconditionally
secure.  We indicateed how a Type 1 protocol, {\bf
QBC1}, and a Type 2 protocol, {\bf
QBC2A}, may be proved secure.  The situation is yet
undecided for Type 3 protocols.  There is no impossibility proof, but
there is no protocol which is clearly secure either. A general
theory of quantum statistical games needs to be developed for
addressing many such QBC
problems in a satisfactory manner.

The content of this paper hopefully makes clear the vast
richness of this subject yet to be uncovered, especially for protocols
that can be practically implemented in a realistic environment.

\newpage
\section*{Appendix A: no impossibility theorem without QBC
definition}
\addcontentsline{toc}{section}{Appendix A:  no impossibility theorem
without QBC definition}
\setcounter{equation}{0}

It is generally believed by mathematicians that a mathematical theorem
can only be obtained from precise mathematical definitions.  In the
impossibility proof of trisecting the angle $\pi/3$ by straightedge
and compass only, for example, the action of these two instruments is precisely
captured mathematically by a quadratic extension field of the rational
numbers.  Do we need a definition of a QBC protocol to have a theorem
which says that unconditionally secure QBC is impossible?  After all, E. Witten
got a Fields Medal in mathematics for work that made essential use of the Feynman
path integral that  M. Atiyah, a former Fields medalist and judge on
the medal's decision panel, commented:  ``... provided one believes
that the integral makes sense,'' to which Witten had the reply: ''We
have forty years of experience of computing these types of integrals''
\cite{schmid}.  Regardless of one's opinion concerning the Feynman
path integral (which, I think, is one of the greatest scientific
creations), it is not similar to a QBC protocol which, unlike the path
integral, has no definite
expression that could serve as a starting point.

A closer analogy to a QBC protocol is an ``effectively computable''
function, a function whose value for any specific argument can be
``mechanically'' obtained in a finite number of steps without the
intervention of ``intelligence.''  The well-known Church-Turing thesis
says that any effectively computable function can be computed
recursively or by a Turing machine. It can be cast as an impossibility
statement: there is no effective procedure that cannot be simulated by
a Turing machine. It was found that a function that
can be computed by a method that is clearly effective, such as Post
machines and Markov algorithms, is indeed also Turing-computable.
However, nobody calls the Church-Turing thesis the Church-Turing
theorem.  This is because there is no mathematical definition of an
effective procedure.  The logical possibility is open that someday a
procedure is found that is intuitively or even physically effective,
but which can compute a nonrecursive arithmetical function.

Thus, in the absence of a precise definition of a QBC protocol, one
would have at best an ``impossibility thesis,'' not an impossibility
{\em theorem}.  (This view was emphasized to the author by Masanao
Ozawa.)  This concern about definition is not scholasticism.  There is no
definition that would characterize all classical cryptographic
protocols, say for bit commitment, partly because, I believe, of the
open possibilities described in Section \ref{sec:newprot} of this
paper.  It is at least not clear why a definition in the more general
quantum case can ever be found. Just as
there appear to be many different forms of effective procedures, there are
many different QBC protocol types that appear not to be captured by
the impossibility proof formulation.  To uphold just the
``impossibility thesis,'' one would need to prove that unconditionally
secure QBC is impossible in each of these types --- four of them are
given in this paper.  My contention is that not only is there no
impossibility proof for these four types, but in fact unconditional
security can be obtained in at least three of them.

\newpage
\section*{Appendix B:  unknown versus random parameters and ``Church
of the Larger Hilbert Space''}
\addcontentsline{toc}{section}{Appendix B:  unknown versus random
parameters and ``Church of the Larger Hilbert Space''}
\renewcommand\theequation{B.\arabic{equation}}
\setcounter{equation}{0}

A considerable amount of confusion surrounds the equivalence between
the use of classical random numbers and their quantum entanglement
purification via the doctrine of ``Church of the Larger Hilbert
Space,'' which is employed in various subareas of quantum information
and quantum cryptography.  There is also confusion about whether there
exists a secret parameter with no probability distribution that can
nevertheless be automatized by a machine.  These questions are tangled
up with a basic assumption or assertion of the impossibility proof
that, in any QBC protocol, there is a publicly known pure state
$\ket{\Phi}$ to start with, which results in a publicly known
$\ket{\Phi_\sb}$ at the end of commitment, connected to $\ket{\Phi}$ by a publicly
known unitary transformation.  In this Appendix, we will show that in
the making of a QBC cryptographic system, some external agent is
always involved, and the system is always open; thus, the above
assertion is untenable.  In the process, we hope to bring out some
clear demarcations that would dispel various confusions.

To begin with, not every unknown parameter can be, or should be,
modelled as a random variable for different reasons, which is
well-known in classical statistics.  One reason is the impossibility
of assigning probabilities to an infinite sample space in some
situations, such as a uniformly distributed random variable with the
values in the positive integers ${\mathbb N}$ or the real numbers
${\mathbb R}$, and similarly on general countably infinite or
uncountable spaces.  This
situation occurs in {\bf QBC4} of Section \ref{sec:type4}, when the
space of all possible actions is of arbitrarily large cardinality, or,
say, just $| {\mathbb N} |$, even though everything is finite to
start with.  A second reason is that there may be no meaningful
ensemble for the parameter $r$, which should be just left as an
unknown parameter to be drawn from a given set, finite or infinite.
This happens in various circumstances, such as the measurement of a
physical (say, astronomical) characteristic that takes on a fixed
value to be estimated.  Such estimation of an unknown parameter
without the use of a priori information on its distribution is,
in fact, very common.  In quantum teleportation, one talks about the
fidelity of receiving a state $\ket{\psi}$ of a qubit that is just
unknown, not with respect to any uniform distribution, so that an
ensemble is described by the density operator $I/2$.  The ensemble is,
rather, $\ket{\psi},\ket{\psi},\ldots$, and the scheme is supposed to
work for any $\ket{\psi}$, not on $I/2$.  A third reason is that often
$r$ is subjected to the control and decision of an agent, and no
probability distribution has a meaning in terms of relative frequency,
as it may have in other cases.  Indeed, the frequency
interpretation of probability has been rejected from the beginning in
decision theory \cite{fine}, for applications to which the
``subjective'' interpretations are more meaningful.  Not only may $r$
be used only once, but also the controlling agent may use it
repeatedly (i.e., $r,r,\ldots$) in an {\it actual} ensemble once it is
decided upon.  There is no actual ensemble that yields
$r_1,r_2,\ldots$ according to whatever probability distribution.

This last situation happens in the use of anonymous states in a QBC
protocol.  Suppose Babe generates (\ref{eq:purif}) instead of
$\ket{\psi_k}$ in an entanglement purification.  The state of
$\cH^{B_1}$ or $\cH^{B_1} \otimes \cH^{B_2}$ is {\em anonymous} to
Adam because he does not know exactly what it is --- Babe has the
freedom to choose $\lambda_k$ in (\ref{eq:purif}).  According to the
Secrecy Principle of Section \ref{sec:newprot}, under a proper
concealing condition she can pick $\{\lambda_k\}$ with any rule made
up by her and unknown to Adam, either for one use or in repeated uses
of (\ref{eq:purif}) for a sequence of different bit commitments.  In
such a sequence, she can use exactly the same value, or use different
values generated according to the parameters decided by another
value.  For example, assuming all possible $\{\lambda_k\}$ form a
finite set, she can pick one randomly and stick to it in a sequence of
commitments.  This can clearly be automatized, and the result does not
appear to Adam as an ensemble with a distribution $\{\lambda_k\}$, but
rather as an ``ensemble'' with one fixed unknown $\{\lambda_k\}$.

Now we are led to the equivalence between random numbers and their
quantum purifications.  I assert that the following sequence of states
generated by random numbers with probability $\{\lambda_k\}$ in
$\cH^{B_1}$,
\begin{equation}
\phi_1,\ldots,\phi_l,\ldots \qquad \ket{\phi_l} \in \{ \ket{\psi_k}\}
\label{eq:rseq}
\end{equation}
is {\em different} from the $\cH^{B_1}$ states in the sequence
obtained by the purification $\Psi$ of (\ref{eq:purif}):
\begin{equation}
\Phi_1,\ldots,\Phi_l,\ldots \qquad \ket{\Phi} = \sum_k
\sqrt{\lambda_k} \ket{\psi_k} \ket{f_k},
\label{eq:pseq}
\end{equation}
where $\ket{\Phi_l} = \ket{\Phi}$ for every $l$.  In a specific
instance $\phi_l$ of (\ref{eq:rseq}) there is no average over $\{
\ket{\psi_k}\}$, but there is always such an average for each $\Phi_l$. Under (\ref{eq:pseq}), which can be used for cheating in the form
(\ref{eq:entstate}), the agent controlling $\cH^{B_2}$ of
(\ref{eq:purif}) or $\cH^A$ of (\ref{eq:entstate}) can select a
preferred ensemble in $\cH^{B_1}$ or $\cH^B$, which makes EPR cheating
possible.  On the other hand, the ensemble in (\ref{eq:rseq}) is fixed
and cannot be changed.  Quantum entanglement is a physical resource
that needs to be established. Not all randomness is reducible to that
of quantum entanglement\footnote[2]{Note that even if it is, an agent can
cheat only if he controls $\cH^{B_2}$ in (\ref{eq:purif}).}.  Indeed,
(\ref{eq:rseq}) does not allow EPR cheating.  This is the situation in
{\bf QBC4}, created through Babe's questioning on Adam's measurement
purification states.  Note that the agent controlling (\ref{eq:rseq})
or (\ref{eq:pseq}) may choose to generate any $\psi_k$ on the
$\phi_l$'s or $\Phi_l$'s.  This situation with unknown
parameters is also relevant
to our Type 3 protocols.

The above difference can be rephrased as follows.  In the density
operator expansion
\begin{equation}
\rho = \sum_k \lambda_k \ketbra{\psi_k}{\psi_k},
\label{eq:doexp}
\end{equation}
the randomness in $k$ may come from a variety of sources.  If all of
it comes from quantum entanglement, then (\ref{eq:purif}) applies,
and the agent controlling $\cH^{B_2}$ can select the ensemble in
$\cH^{B_1}$.  If some of it comes from elsewhere, it would not be
equivalent to (\ref{eq:purif}), and ensemble selection or entanglement
cheating is limited or becomes impossible, depending on the exact form
of the joint state.  The occurrence of such non-entanglement
randomness is always possible because the system is subject to
intervention by agents.  In any meaningful and realistic
formulation of the problem, the agents' possible actions are
infinitely varying and open.  They cannot be described as a public
$\ket{\Phi}$ being transformed in a closed system to another public
$\ket{\Phi_\sb}$.  Indeed, for QBC there is in general a
game-theoretic situation, where both parties can choose actions
unknown to the other party.

\newpage
\section*{Appendix C:  protocol QBC4}
\addcontentsline{toc}{section}{Appendix C:  protocol QBC4}
\renewcommand\theequation{C.\arabic{equation}}
\setcounter{equation}{0}

Here we fill in certain mathematical details on {\bf QBC4}.  We
consider first the case when Babe asks no question on the evidence.

Adam can guarantee concealing by using uniform probability $1/2^{n-1}$
for each sequence of either parity.  In that case, $\rho^B_0 -
\rho^B_1$ factorizes into products of individual qubit parts.  Let
${\bf j} = \{j_1,\ldots,j_n\} \in \{0,1\}^n$, $P_{l0} =
\ketbra{\phi}{\phi}$, $P_{l1} = \ketbra{\phi'}{\phi'}$, $l \in
\{1,\ldots,n\}$.  Let $\Lambda_0 = \{ {\bf j} | \bigoplus^n_{l=1} j_l
= 0\}$, $\Lambda_1 = \{ {\bf j}| \bigoplus^n_{l=1}j_l=1\}$ be the
even- and odd-parity $n$-bit sets.  Then
\begin{equation}
\rho^B_\sb = \frac{1}{2^{n-1}} \sum_{{\bf j} \in \Lambda_\sb}
\bigotimes^n_{l=1} P_{lj_l},\qquad \sb \in \{0,1\}\,,
\end{equation}
and so
\begin{equation}
\rho^B_0 - \rho^B_1 = \frac{1}{2^{n-1}} \bigotimes^n_{l=1}
\left(P_{l0}-P_{l1}\right)\,.
\end{equation}
Thus, Babe's optimum quantum decision reduces to optimally
discriminating between $\ket{\phi}$ and $\ket{\phi'}$ for each qubit
individually, and then seeing whether there is an even or odd number
of $\ket{\phi'}$'s.  the optimum error probability $p_e$ for each
qubit is well-known \cite{hel,yue2},
\begin{equation}
p_e = \frac{1}{2}-\frac{1}{2}\sqrt{1-|\braket{\phi}{\phi'}|^2}.
\label{eq:pe}
\end{equation}
The optimum error probability $\bar{P}^B_c$ of correct bit decision on
the sequence is, from the even and odd binomial sums, given by
\begin{equation}
\bar{P}^B_c = \frac{1}{2} + \frac{1}{2}(1-2p_e)^n\,.
\label{eq:perror}
\end{equation}
Thus $\bar{P}^B_c$ is close to 1/2 exponentially in $n$ independently
of $1/2 \ge p_e > 0$.  

After committing $\ket{\Phi_0}$, Adam can still try to cheat with the
$\{ \ket{e_i} \}$ measurement by declaring one qubit to be in a state
different from the actual one.  the probability of success is $P^A_c =
|\braket{\phi}{\phi'}|^2 \equiv \epsilon_1$, a design parameter of the
protocol. It can be made $\epsilon$-concealing by choosing
\begin{equation}
| \braket{\phi}{\phi'} |^2 \equiv \epsilon_1 \le \epsilon
\end{equation}
and, from (\ref{eq:perror}), choosing $n_0$ to satisfy
\begin{equation}
(1-\epsilon_1)^{n_0} \le 4\epsilon^2\,.
\label{eq:epsis}
\end{equation}

When Adam presents the additional $\ket{i_l}$'s in $S_0$ or other
sets, Babe's density operator $\rho^B_0$ is diagonal, similar to
(\ref{eq:pseq}), in the basis that diagonalizes each pair $\cH_{l2}
\otimes \cK_{l2}$.  Her optimum decision reduces to optimally
discriminating between the two density operators corresponding to $i_l
= 0$ and $i_l = 1$ for each of these $n$ pairs, and then choosing the
total resulting parity from the $n$ decisions.  Thus, $\bar{P}^B_c$ is
given by (\ref{eq:perror}) with $p_e$ given by the optimum pair
decision, which just yields a different function of $\epsilon_1$ from
(\ref{eq:pe}).  For any fixed $\epsilon_1$, $\bar{P}^B_c$ can be made
smaller than any $\epsilon$ as in (\ref{eq:epsis}) with a large enough
$n_0$.

The following theorem characterizes Adam's optimal probability of
cheating $\bar{P}^A_c$ when $\ket{\Phi_0}$ of (\ref{eq:entstate}) is
used with resulting $\rho^B_\sb$ and $F$ between them.

\noindent{{\bf Theorem}}
\begin{equation}
F^2 \le \bar{P}^A_c \le F.
\label{eq:fid}
\end{equation}
The bounds (\ref{eq:fid}) are identical to (21) for $\tilde{P}^A_c$ in
Ref.~\cite{yue4}.  It can be seen from Appendix A, (6), and (18) of
\cite{yue4} that actually $\tilde{P}^A_c = \bar{P}^A_c$.

The following theorem \cite{yue2} is also used in Section \ref{sec:type4}.

\noindent{{\bf Theorem} (local state invariance).\,\, Let $\rho^{AB}$
  be a state on $\cH^A \otimes \cH^B$ with $\rho^B
  \equiv \tr_A \rho^{AB}$.  The state $\rho^B$
  remains invariant under any quantum operation on $\cH^A$ alone.}
\vskip 12pt

\newpage
\section*{Appendix D:  physical limits and unconditional security}
\addcontentsline{toc}{section}{Appendix D:  physical limits and
unconditional security}
\renewcommand\theequation{D.\arabic{equation}}
\setcounter{equation}{0}

In cryptography, a system is typcally called unconditionally secure if it
cannot be broken with infinite computational power, i.e., its security
is not based on computational complexity of any kind.  In quantum
cryptography, the system's security depends on the validity of the laws
of quantum physics and not on the limits of computational power, so this
security is unconditional.  More
broadly, one can say that in {\em physical cryptography}, the system's
security is based on facts of our physical world which are immutable,
and hence is unconditional also.  Indeed, the laws of physics are part of the {\em facts of Nature}, which include both the
laws and the initial conditions of the Universe that give rise to the
world we live in.  For example, we can
exploit and utilize the background radiation from the sun or even the
cosmos, because they are always there, not removable by any
technological advance.  Such physical limits are fundamentally
different from ones that arise from computational complexity, quantum
or classical.

Similarly, there are facts of nature that impose physical limits on
the possible number of qubits one may use in entanglement.  What may
be surprising is that the number so limited is small on just an exponential scale.  By
various estimates, the total number of elementary fermions in the
world is ${< \atop \sim} 10^{89} < 2^{400}$ \cite{pea}.  If $E_1$ is the energy
range available with separation $\Delta E \sim \hbar/\Delta t$, taking
$\Delta t$ to be the age of the Universe ($< 2^{40}$ sec.), the total number of qubits
available with energy $E_1$ is $2^{400} \log \frac{E_1 \Delta
t}{\hbar} {< \atop \sim} 2^{410}$. For the boson electromagnetic field with total energy
$E_2$, the number of qubits available is $\Delta t \log
\frac{E_2}{\Delta \tau\hbar}$ from the bit capacity of a boson field
\cite{yo,cd}. Taking $\tau$ to be the Planck time $10^{-44}$ sec and
$E_2$ the total electromagnetic radiation energy in the Universe
\cite{pea}, this yields $< 2^{380}$ qubits.  Thus, one can entangle no
more than $\sim 2^{410}$ binary possibilities.

\newpage
\section*{Acknowledgment}
\addcontentsline{toc}{section}{Acknowledgment}

I would like to thank G.M. D'Ariano and M. Ozawa for useful
discussions, and C. Bennett, I. Chuang, C. Cr\'epeau, D. Gottesman and
T. Rudolph for their comments. This work was supported by the Defense Advanced Research
Projects Agency and the Army Research Office.

\end{document}